 \documentclass[conference,10pt,twoside,letterpaper]{IEEEtran}

\usepackage{mleftright}
\usepackage{cite}
\usepackage[T1]{fontenc}
\usepackage{graphicx}
\usepackage{mathtools}
\usepackage{amssymb}
\usepackage{amsmath}
\usepackage{paralist}
\usepackage{subfigure}
\usepackage{booktabs} 
\usepackage{algorithm}
\usepackage{algorithmic}
\usepackage{amsthm}
\usepackage{multirow}
\usepackage{microtype}
\usepackage{balance}

\usepackage{amssymb}
\usepackage{amsfonts}
\usepackage{mathrsfs}
\usepackage{xspace}
\usepackage{bm}
\usepackage{upgreek}

\newcommand{\safemath}[2]{\newcommand{#1}{\ensuremath{#2}\xspace}}

\safemath{\bma}{\mathbf{a}}
\safemath{\bmb}{\mathbf{b}}
\safemath{\bmc}{\mathbf{c}}
\safemath{\bmd}{\mathbf{d}}
\safemath{\bme}{\mathbf{e}}
\safemath{\bmf}{\mathbf{f}}
\safemath{\bmg}{\mathbf{g}}
\safemath{\bmh}{\mathbf{h}}
\safemath{\bmi}{\mathbf{i}}
\safemath{\bmj}{\mathbf{j}}
\safemath{\bmk}{\mathbf{k}}
\safemath{\bml}{\mathbf{l}}
\safemath{\bmm}{\mathbf{m}}
\safemath{\bmn}{\mathbf{n}}
\safemath{\bmo}{\mathbf{o}}
\safemath{\bmp}{\mathbf{p}}
\safemath{\bmq}{\mathbf{q}}
\safemath{\bmr}{\mathbf{r}}
\safemath{\bms}{\mathbf{s}}
\safemath{\bmt}{\mathbf{t}}
\safemath{\bmu}{\mathbf{u}}
\safemath{\bmv}{\mathbf{v}}
\safemath{\bmw}{\mathbf{w}}
\safemath{\bmx}{\mathbf{x}}
\safemath{\bmy}{\mathbf{y}}
\safemath{\bmz}{\mathbf{z}}
\safemath{\bmzero}{\mathbf{0}}
\safemath{\bmone}{\mathbf{1}}

\bmdefine{\biad}{a}
\bmdefine{\bibd}{b}
\bmdefine{\bicd}{c}
\bmdefine{\bidd}{d}
\bmdefine{\bied}{e}
\bmdefine{\bifd}{f}
\bmdefine{\bigd}{g}
\bmdefine{\bihd}{h}
\bmdefine{\biid}{i}
\bmdefine{\bijd}{j}
\bmdefine{\bikd}{k}
\bmdefine{\bild}{l}
\bmdefine{\bimd}{m}
\bmdefine{\bind}{n}
\bmdefine{\biod}{o}
\bmdefine{\bipd}{p}
\bmdefine{\biqd}{q}
\bmdefine{\bird}{r}
\bmdefine{\bisd}{s}
\bmdefine{\bitd}{t}
\bmdefine{\biud}{u}
\bmdefine{\bivd}{v}
\bmdefine{\biwd}{w}
\bmdefine{\bixd}{x}
\bmdefine{\biyd}{y}
\bmdefine{\bizd}{z}

\bmdefine{\bixid}{\xi}
\bmdefine{\bilambdad}{\lambda}
\bmdefine{\bimud}{\mu}
\bmdefine{\bithetad}{\theta}
\bmdefine{\biphid}{\phi}
\bmdefine{\bideltad}{\delta}

\safemath{\bmia}{\biad}
\safemath{\bmib}{\bibd}
\safemath{\bmic}{\bicd}
\safemath{\bmid}{\bidd}
\safemath{\bmie}{\bied}
\safemath{\bmif}{\bifd}
\safemath{\bmig}{\bigd}
\safemath{\bmih}{\bihd}
\safemath{\bmii}{\biid}
\safemath{\bmij}{\bijd}
\safemath{\bmik}{\bikd}
\safemath{\bmil}{\bild}
\safemath{\bmim}{\bimd}
\safemath{\bmin}{\bind}
\safemath{\bmio}{\biod}
\safemath{\bmip}{\bipd}
\safemath{\bmiq}{\biqd}
\safemath{\bmir}{\bird}
\safemath{\bmis}{\bisd}
\safemath{\bmit}{\bitd}
\safemath{\bmiu}{\biud}
\safemath{\bmiv}{\bivd}
\safemath{\bmiw}{\biwd}
\safemath{\bmix}{\bixd}
\safemath{\bmiy}{\biyd}
\safemath{\bmiz}{\bizd}

\safemath{\bmxi}{\bixid}
\safemath{\bmlambda}{\bilambdad}
\safemath{\bmmu}{\bimud}
\safemath{\bmtheta}{\bithetad}
\safemath{\bmphi}{\biphid}
\safemath{\bmdelta}{\bideltad}

\safemath{\bA}{\mathbf{A}}
\safemath{\bB}{\mathbf{B}}
\safemath{\bC}{\mathbf{C}}
\safemath{\bD}{\mathbf{D}}
\safemath{\bE}{\mathbf{E}}
\safemath{\bF}{\mathbf{F}}
\safemath{\bG}{\mathbf{G}}
\safemath{\bH}{\mathbf{H}}
\safemath{\bI}{\mathbf{I}}
\safemath{\bJ}{\mathbf{J}}
\safemath{\bK}{\mathbf{K}}
\safemath{\bL}{\mathbf{L}}
\safemath{\bM}{\mathbf{M}}
\safemath{\bN}{\mathbf{N}}
\safemath{\bO}{\mathbf{O}}
\safemath{\bP}{\mathbf{P}}
\safemath{\bQ}{\mathbf{Q}}
\safemath{\bR}{\mathbf{R}}
\safemath{\bS}{\mathbf{S}}
\safemath{\bT}{\mathbf{T}}
\safemath{\bU}{\mathbf{U}}
\safemath{\bV}{\mathbf{V}}
\safemath{\bW}{\mathbf{W}}
\safemath{\bX}{\mathbf{X}}
\safemath{\bY}{\mathbf{Y}}
\safemath{\bZ}{\mathbf{Z}}

\safemath{\bZero}{\mathbf{0}}
\safemath{\bOne}{\mathbf{1}}
\safemath{\bDelta}{\mathbf{\Delta}}
\safemath{\bLambda}{\mathbf{\UpLambda}}
\safemath{\bPhi}{\mathbf{\Upphi}}
\safemath{\bSigma}{\mathbf{\Upsigma}}
\safemath{\bOmega}{\mathbf{\Upomega}}
\safemath{\bTheta}{\mathbf{\Uptheta}}

\bmdefine{\biAd}{A}
\bmdefine{\biBd}{B}
\bmdefine{\biCd}{C}
\bmdefine{\biDd}{D}
\bmdefine{\biEd}{E}
\bmdefine{\biFd}{F}
\bmdefine{\biGd}{G}
\bmdefine{\biHd}{H}
\bmdefine{\biId}{I}
\bmdefine{\biJd}{J}
\bmdefine{\biKd}{K}
\bmdefine{\biLd}{L}
\bmdefine{\biMd}{M}
\bmdefine{\biOd}{N}
\bmdefine{\biPd}{O}
\bmdefine{\biQd}{P}
\bmdefine{\biRd}{R}
\bmdefine{\biSd}{S}
\bmdefine{\biTd}{T}
\bmdefine{\biUd}{U}
\bmdefine{\biVd}{V}
\bmdefine{\biWd}{W}
\bmdefine{\biXd}{X}
\bmdefine{\biYd}{Y}
\bmdefine{\biZd}{Z}

\bmdefine{\biDelta}{\Delta}
\bmdefine{\biLambda}{\Lambda}
\bmdefine{\biPhi}{\Phi}
\bmdefine{\biSigma}{\Sigma}
\bmdefine{\biOmega}{\Omega}
\bmdefine{\biTheta}{\Theta}

\safemath{\bimA}{\biAd}
\safemath{\bimB}{\biBd}
\safemath{\bimC}{\biCd}
\safemath{\bimD}{\biDd}
\safemath{\bimE}{\biEd}
\safemath{\bimF}{\biFd}
\safemath{\bimG}{\biGd}
\safemath{\bimH}{\biHd}
\safemath{\bimI}{\biId}
\safemath{\bimJ}{\biJd}
\safemath{\bimK}{\biKd}
\safemath{\bimL}{\biLd}
\safemath{\bimM}{\biMd}
\safemath{\bimN}{\biNd}
\safemath{\bimO}{\biOd}
\safemath{\bimP}{\biPd}
\safemath{\bimQ}{\biQd}
\safemath{\bimR}{\biRd}
\safemath{\bimS}{\biSd}
\safemath{\bimT}{\biTd}
\safemath{\bimU}{\biUd}
\safemath{\bimV}{\biVd}
\safemath{\bimW}{\biWd}
\safemath{\bimX}{\biXd}
\safemath{\bimY}{\biYd}
\safemath{\bimZ}{\biZd}

\safemath{\bimDelta}{\biDelta}
\safemath{\bimLambda}{\biLambda}
\safemath{\bimPhi}{\biPhi}
\safemath{\bimSigma}{\biSigma}
\safemath{\bimOmega}{\biOmega}
\safemath{\bimTheta}{\biTheta}

\safemath{\setA}{\mathcal{A}}
\safemath{\setB}{\mathcal{B}}
\safemath{\setC}{\mathcal{C}}
\safemath{\setD}{\mathcal{D}}
\safemath{\setE}{\mathcal{E}}
\safemath{\setF}{\mathcal{F}}
\safemath{\setG}{\mathcal{G}}
\safemath{\setH}{\mathcal{H}}
\safemath{\setI}{\mathcal{I}}
\safemath{\setJ}{\mathcal{J}}
\safemath{\setK}{\mathcal{K}}
\safemath{\setL}{\mathcal{L}}
\safemath{\setM}{\mathcal{M}}
\safemath{\setN}{\mathcal{N}}
\safemath{\setO}{\mathcal{O}}
\safemath{\setP}{\mathcal{P}}
\safemath{\setQ}{\mathcal{Q}}
\safemath{\setR}{\mathcal{R}}
\safemath{\setS}{\mathcal{S}}
\safemath{\setT}{\mathcal{T}}
\safemath{\setU}{\mathcal{U}}
\safemath{\setV}{\mathcal{V}}
\safemath{\setW}{\mathcal{W}}
\safemath{\setX}{\mathcal{X}}
\safemath{\setY}{\mathcal{Y}}
\safemath{\setZ}{\mathcal{Z}}
\safemath{\emptySet}{\varnothing}

\safemath{\colA}{\mathscr{A}}
\safemath{\colB}{\mathscr{B}}
\safemath{\colC}{\mathscr{C}}
\safemath{\colD}{\mathscr{D}}
\safemath{\colE}{\mathscr{E}}
\safemath{\colF}{\mathscr{F}}
\safemath{\colG}{\mathscr{G}}
\safemath{\colH}{\mathscr{H}}
\safemath{\colI}{\mathscr{I}}
\safemath{\colJ}{\mathscr{J}}
\safemath{\colK}{\mathscr{K}}
\safemath{\colL}{\mathscr{L}}
\safemath{\colM}{\mathscr{M}}
\safemath{\colN}{\mathscr{N}}
\safemath{\colO}{\mathscr{O}}
\safemath{\colP}{\mathscr{P}}
\safemath{\colQ}{\mathscr{Q}}
\safemath{\colR}{\mathscr{R}}
\safemath{\colS}{\mathscr{S}}
\safemath{\colT}{\mathscr{T}}
\safemath{\colU}{\mathscr{U}}
\safemath{\colV}{\mathscr{V}}
\safemath{\colW}{\mathscr{W}}
\safemath{\colX}{\mathscr{X}}
\safemath{\colY}{\mathscr{Y}}
\safemath{\colZ}{\mathscr{Z}}

\safemath{\opA}{\mathbb{A}}
\safemath{\opB}{\mathbb{B}}
\safemath{\opC}{\mathbb{C}}
\safemath{\opD}{\mathbb{D}}
\safemath{\opE}{\mathbb{E}}
\safemath{\opF}{\mathbb{F}}
\safemath{\opG}{\mathbb{G}}
\safemath{\opH}{\mathbb{H}}
\safemath{\opI}{\mathbb{I}}
\safemath{\opJ}{\mathbb{J}}
\safemath{\opK}{\mathbb{K}}
\safemath{\opL}{\mathbb{L}}
\safemath{\opM}{\mathbb{M}}
\safemath{\opN}{\mathbb{N}}
\safemath{\opO}{\mathbb{O}}
\safemath{\opP}{\mathbb{P}}
\safemath{\opQ}{\mathbb{Q}}
\safemath{\opR}{\mathbb{R}}
\safemath{\opS}{\mathbb{S}}
\safemath{\opT}{\mathbb{T}}
\safemath{\opU}{\mathbb{U}}
\safemath{\opV}{\mathbb{V}}
\safemath{\opW}{\mathbb{W}}
\safemath{\opX}{\mathbb{X}}
\safemath{\opY}{\mathbb{Y}}
\safemath{\opZ}{\mathbb{Z}}
\safemath{\opZero}{\mathbb{O}}
\safemath{\identityop}{\opI}

\safemath{\veca}{\bma}
\safemath{\vecb}{\bmb}
\safemath{\vecc}{\bmc}
\safemath{\vecd}{\bmd}
\safemath{\vece}{\bme}
\safemath{\vecf}{\bmf}
\safemath{\vecg}{\bmg}
\safemath{\vech}{\bmh}
\safemath{\veci}{\bmi}
\safemath{\vecj}{\bmj}
\safemath{\veck}{\bmk}
\safemath{\vecl}{\bml}
\safemath{\vecm}{\bmm}
\safemath{\vecn}{\bmn}
\safemath{\veco}{\bmo}
\safemath{\vecp}{\bmp}
\safemath{\vecq}{\bmq}
\safemath{\vecr}{\bmr}
\safemath{\vecs}{\bms}
\safemath{\vect}{\bmt}
\safemath{\vecu}{\bmu}
\safemath{\vecv}{\bmv}
\safemath{\vecw}{\bmw}
\safemath{\vecx}{\bmx}
\safemath{\vecy}{\bmy}
\safemath{\vecz}{\bmz}

\safemath{\veczero}{\bmzero}
\safemath{\vecone}{\bmone}
\safemath{\vecxi}{\bmxi}
\safemath{\veclambda}{\bmlambda}
\safemath{\vecmu}{\bmmu}
\safemath{\vectheta}{\bmtheta}
\safemath{\vecphi}{\bmphi}
\safemath{\vecdelta}{\bmdelta}

\safemath{\matA}{\bA}
\safemath{\matB}{\bB}
\safemath{\matC}{\bC}
\safemath{\matD}{\bD}
\safemath{\matE}{\bE}
\safemath{\matF}{\bF}
\safemath{\matG}{\bG}
\safemath{\matH}{\bH}
\safemath{\matI}{\bI}
\safemath{\matJ}{\bJ}
\safemath{\matK}{\bK}
\safemath{\matL}{\bL}
\safemath{\matM}{\bM}
\safemath{\matN}{\bN}
\safemath{\matO}{\bO}
\safemath{\matP}{\bP}
\safemath{\matQ}{\bQ}
\safemath{\matR}{\bR}
\safemath{\matS}{\bS}
\safemath{\matT}{\bT}
\safemath{\matU}{\bU}
\safemath{\matV}{\bV}
\safemath{\matW}{\bW}
\safemath{\matX}{\bX}
\safemath{\matY}{\bY}
\safemath{\matZ}{\bZ}
\safemath{\matzero}{\bmzero}

\safemath{\matDelta}{\bDelta}
\safemath{\matLambda}{\bLambda}
\safemath{\matPhi}{\bPhi}
\safemath{\matSigma}{\bSigma}
\safemath{\matOmega}{\bOmega}
\safemath{\matTheta}{\bTheta}

\safemath{\matidentity}{\matI}
\safemath{\matone}{\matO}

\safemath{\rnda}{A}
\safemath{\rndb}{B}
\safemath{\rndc}{C}
\safemath{\rndd}{D}
\safemath{\rnde}{E}
\safemath{\rndf}{F}
\safemath{\rndg}{G}
\safemath{\rndh}{H}
\safemath{\rndi}{I}
\safemath{\rndj}{J}
\safemath{\rndk}{K}
\safemath{\rndl}{L}
\safemath{\rndm}{M}
\safemath{\rndn}{N}
\safemath{\rndo}{O}
\safemath{\rndp}{P}
\safemath{\rndq}{Q}
\safemath{\rndr}{R}
\safemath{\rnds}{S}
\safemath{\rndt}{T}
\safemath{\rndu}{U}
\safemath{\rndv}{V}
\safemath{\rndw}{W}
\safemath{\rndx}{X}
\safemath{\rndy}{Y}
\safemath{\rndz}{Z}

\safemath{\rveca}{\bimA}
\safemath{\rvecb}{\bimB}
\safemath{\rvecc}{\bimC}
\safemath{\rvecd}{\bimD}
\safemath{\rvece}{\bimE}
\safemath{\rvecf}{\bimF}
\safemath{\rvecg}{\bimG}
\safemath{\rvech}{\bimH}
\safemath{\rveci}{\bimI}
\safemath{\rvecj}{\bimJ}
\safemath{\rveck}{\bimK}
\safemath{\rvecl}{\bimL}
\safemath{\rvecm}{\bimM}
\safemath{\rvecn}{\bimN}
\safemath{\rveco}{\bomO}
\safemath{\rvecp}{\bimP}
\safemath{\rvecq}{\bimQ}
\safemath{\rvecr}{\bimR}
\safemath{\rvecs}{\bimS}
\safemath{\rvect}{\bimT}
\safemath{\rvecu}{\bimU}
\safemath{\rvecv}{\bimV}
\safemath{\rvecw}{\bimW}
\safemath{\rvecx}{\bimX}
\safemath{\rvecy}{\bimY}
\safemath{\rvecz}{\bimZ}

\safemath{\rvecxi}{\bmxi}
\safemath{\rveclambda}{\bmlambda}
\safemath{\rvecmu}{\bmmu}
\safemath{\rvectheta}{\bmtheta}
\safemath{\rvecphi}{\bmphi}

\safemath{\rmatA}{\bimA}
\safemath{\rmatB}{\bimB}
\safemath{\rmatC}{\bimC}
\safemath{\rmatD}{\bimD}
\safemath{\rmatE}{\bimE}
\safemath{\rmatF}{\bimF}
\safemath{\rmatG}{\bimG}
\safemath{\rmatH}{\bimH}
\safemath{\rmatI}{\bimI}
\safemath{\rmatJ}{\bimJ}
\safemath{\rmatK}{\bimK}
\safemath{\rmatL}{\bimL}
\safemath{\rmatM}{\bimM}
\safemath{\rmatN}{\bimN}
\safemath{\rmatO}{\bimO}
\safemath{\rmatP}{\bimP}
\safemath{\rmatQ}{\bimQ}
\safemath{\rmatR}{\bimR}
\safemath{\rmatS}{\bimS}
\safemath{\rmatT}{\bimT}
\safemath{\rmatU}{\bimU}
\safemath{\rmatV}{\bimV}
\safemath{\rmatW}{\bimW}
\safemath{\rmatX}{\bimX}
\safemath{\rmatY}{\bimY}
\safemath{\rmatZ}{\bimZ}

\safemath{\rmatDelta}{\bimDelta}
\safemath{\rmatLambda}{\bimLambda}
\safemath{\rmatPhi}{\bimPhi}
\safemath{\rmatSigma}{\bimSigma}
\safemath{\rmatOmega}{\bimOmega}
\safemath{\rmatTheta}{\bimTheta}

\usepackage{amssymb}
\usepackage{amsfonts}
\usepackage{mathrsfs}
\usepackage{xspace}
\usepackage{bm}
\usepackage{fancyref}
\usepackage{textcomp}

\usepackage{multirow}
\usepackage{stmaryrd}

\newenvironment{textbmatrix}{	\setlength{\arraycolsep}{2.5pt}%
								\big[\begin{matrix}}{\end{matrix}\big]%
								\raisebox{0.08ex}{\vphantom{M}}}

\def\be{\begin{equation}}
\def\ee{\end{equation}}
\def\een{\nonumber \end{equation}}
\def\mat{\begin{bmatrix}}
\def\emat{\end{bmatrix}}
\def\btm{\begin{textbmatrix}}
\def\etm{\end{textbmatrix}}

\def\ba#1\ea{\begin{align}#1\end{align}}
\def\bas#1\eas{\begin{align*}#1\end{align*}}
\def\bs#1\es{\begin{split}#1\end{split}}
\def\bg#1\eg{\begin{gather}#1\end{gather}}
\def\bml#1\eml{\begin{multline}#1\end{multline}}
\def\bi#1\ei{\begin{itemize}#1\end{itemize}}

\newcommand{\lefto}{\mathopen{}\left}

\DeclareMathOperator{\sign}{sign}			
\DeclareMathOperator*{\argmin}{arg\;min}		
\DeclareMathOperator*{\argmax}{arg\;max}		
\DeclareMathOperator{\Exop}{\opE}			
\DeclareMathOperator{\Varop}{\opV\!\mathrm{ar}} 

\newcommand{\abs}[1]{\lefto\lvert#1\right\rvert}		

\newcommand{\vecnorm}[1]{\lefto\lVert#1\right\rVert}		

\safemath{\dirac}{\delta}					
\safemath{\krond}{\dirac}					

\safemath{\upto}{\uparrow}
\safemath{\downto}{\downarrow}
\safemath{\iu}{j}							
\safemath{\ev}{\lambda}						
\safemath{\hilseqspace}{l^{2}}				
\newcommand{\banachfunspace}[1]{\setL^{#1}}	
\safemath{\hilfunspace}{\banachfunspace{2}}	

\safemath{\SNR}{\textit{SNR}} 				
\safemath{\PAR}{\textit{PAR}} 				
\safemath{\No}{N_0}							
\safemath{\Es}{E_s}							
\safemath{\Eb}{E_b}							
\safemath{\EbNo}{\frac{\Eb}{\No}}
\safemath{\EsNo}{\frac{\Es}{\No}}

\DeclareMathOperator{\CHop}{\ensuremath{\opH}} 
\safemath{\tvir}{\rndh_{\CHop}}				
\safemath{\tvtf}{\rndl_{\CHop}}				
\safemath{\spf}{\rnds_{\CHop}}				
\safemath{\bff}{H_{\CHop}}					

\safemath{\ircf}{r_{h}}						
\safemath{\tftvcf}{r_{s}}					
\safemath{\tfcf}{r_{l}}						
\safemath{\bfcf}{r_{H}}						

\safemath{\tcorr}{c_h}						
\safemath{\scf}{c_{s}}						
\safemath{\tfcorr}{c_{l}}					
\safemath{\fcorr}{c_{H}}						

\safemath{\mi}{I}							
\safemath{\capacity}{C}						

\safemath{\normal}{\mathcal{N}}			
\safemath{\jpg}{\mathcal{CN}}			
\safemath{\mchain}{\leftrightarrow}		

\safemath{\dB}{\,\mathrm{dB}}
\safemath{\dBm}{\,\mathrm{dBm}}
\safemath{\Hz}{\,\mathrm{Hz}}
\safemath{\kHz}{\,\mathrm{kHz}}
\safemath{\MHz}{\,\mathrm{MHz}}
\safemath{\GHz}{\,\mathrm{GHz}}
\safemath{\s}{\,\mathrm{s}}
\safemath{\ms}{\,\mathrm{ms}}
\safemath{\mus}{\,\mathrm{\text{\textmu}s}}
\safemath{\ns}{\,\mathrm{ns}}
\safemath{\ps}{\,\mathrm{ps}}
\safemath{\meter}{\,\mathrm{m}}
\safemath{\mm}{\,\mathrm{mm}}
\safemath{\cm}{\,\mathrm{cm}}
\safemath{\m}{\,\mathrm{m}}
\safemath{\W}{\,\mathrm{W}}
\safemath{\mW}{\, \mathrm{mW}}
\safemath{\J}{\,\mathrm{J}}
\safemath{\K}{\,\mathrm{K}}
\safemath{\bit}{\,\mathrm{bit}}
\safemath{\nat}{\,\mathrm{nat}}

\safemath{\define}{\triangleq}			

\safemath{\equivalent}{\sim}
\safemath{\distas}{\sim}					
\safemath{\sdiff}{\Delta}				

\safemath{\reals}{\mathbb{R}}
\safemath{\positivereals}{\reals_{+}}
\safemath{\integers}{\mathbb{Z}}
\safemath{\posint}{\integers_{+}}
\safemath{\naturals}{\mathbb{N}}
\safemath{\posnaturals}{\naturals_{+}}
\safemath{\complexset}{\mathbb{C}}
\safemath{\rationals}{\mathbb{Q}}

\newcommand*{\fancyrefapplabelprefix}{app}		
\newcommand*{\fancyrefthmlabelprefix}{thm}		
\newcommand*{\fancyreflemlabelprefix}{lem}		
\newcommand*{\fancyrefcorlabelprefix}{cor}		
\newcommand*{\fancyrefdeflabelprefix}{def}		
\newcommand*{\fancyrefproplabelprefix}{prop}		
\newcommand*{\fancyrefexmpllabelprefix}{exmpl}
\newcommand*{\fancyrefalglabelprefix}{alg}		
\newcommand*{\fancyreftbllabelprefix}{tbl}		

\frefformat{vario}{\fancyrefseclabelprefix}{Section~#1}
\frefformat{vario}{\fancyrefthmlabelprefix}{Theorem~#1}
\frefformat{vario}{\fancyreftbllabelprefix}{Table~#1}
\frefformat{vario}{\fancyreflemlabelprefix}{Lemma~#1}
\frefformat{vario}{\fancyrefcorlabelprefix}{Corollary~#1}
\frefformat{vario}{\fancyrefdeflabelprefix}{Definition~#1}
\frefformat{vario}{\fancyreffiglabelprefix}{Fig.~#1}
\frefformat{vario}{\fancyrefapplabelprefix}{Appendix~#1}
\frefformat{vario}{\fancyrefeqlabelprefix}{(#1)}
\frefformat{vario}{\fancyrefproplabelprefix}{Proposition~#1}
\frefformat{vario}{\fancyrefexmpllabelprefix}{Example~#1}
\frefformat{vario}{\fancyrefalglabelprefix}{Algorithm~#1}

 \newtheorem{thm}{Theorem}
 \newtheorem{cor}[thm]{Corollary}   
 
 \newtheorem{defi}{Definition}
 \newtheorem{lem}[thm]{Lemma}
 
\safemath{\dictab}{[\,\dicta\,\,\dictb\,]}

\safemath{\ysig}{\bmy}
\safemath{\ysighat}{\hat{\ysig}}
\safemath{\ysigdim}{M}
\safemath{\xsig}{\bmx}
\safemath{\xsigdim}{N}
\safemath{\nx}{n_x}
\safemath{\zsig}{\bmz}
\safemath{\zsigdim}{\ysigdim}
\safemath{\rsig}{\bmr}
\safemath{\Adict}{\bA}
\safemath{\Adicttilde}{\widetilde{\Adict}}
\safemath{\Adictdim}{\outputdim\times\xsigdim}
\safemath{\avec}{\bma}
\safemath{\avectilde}{\tilde{\avec}}
\safemath{\Bdict}{\bB}
\safemath{\Bdicttilde}{\widetilde{\Bdict}}
\safemath{\Cdict}{\bC}
\safemath{\cvec}{\bmc}
\safemath{\Ddict}{\bD}
\safemath{\Ddictdim}{\ysigdim\times\xsigdim}
\safemath{\dvec}{\bmd}
\safemath{\Ddicttilde}{\widetilde{\bD}}
\safemath{\Bonb}{\bB}
\safemath{\bvec}{\bmb}
\safemath{\Bonbdim}{\ysigdim\times\ysigdim}
\safemath{\noise}{\bmn}
\safemath{\noisedim}{\ysigim}
\safemath{\err}{\bme}
\safemath{\errdim}{\ysigdim}
\safemath{\errset}{\setE}
\safemath{\nerr}{n_e}
\safemath{\delop}{\bP_\errset}
\safemath{\delopc}{\bP_{{\errset}^c}}

\safemath{\cplxi}{\imath}
\safemath{\cplxj}{\jmath}

\safemath{\dict}{\matD}
\safemath{\inputdim}{N}		
\safemath{\outputdim}{M}		
\safemath{\sparsity}{S}	
\safemath{\inputdimA}{{N_a}}	
\safemath{\inputdimB}{{N_b}}	
\safemath{\elemA}{{n_a}}	
\safemath{\elemB}{{n_b}}	
\safemath{\resA}{\matR_a}	
\safemath{\resB}{\matR_b}	
\safemath{\subD}{\matS} 
\safemath{\subA}{\matS_a} 
\safemath{\subB}{\matS_b} 
\safemath{\dicta}{\matA} 	
\safemath{\dictb}{\matB} 	
\safemath{\hollowS}{H}
\safemath{\hollowA}{H_a}
\safemath{\hollowB}{H_b}
\safemath{\cross}{Z}
\safemath{\coh}{\mu_d}			
\safemath{\coha}{\mu_a}			
\safemath{\cohb}{\mu_b}			
\safemath{\mubs}{\nu}	
\safemath{\cohm}{\mu_m} 
\safemath{\dictset}{\setD}	
\safemath{\dictsetp}{\dictset(\coh,\coha,\cohb)}	
\safemath{\dictsetgen}{\dictset_\text{gen}}
\safemath{\dictsetgenp}{\dictsetgen(\coh)}
\safemath{\dictsetonb}{\dictset_\text{onb}}
\safemath{\dictsetonbp}{\dictsetonb(\coh)}

\safemath{\leftside}{U}
\safemath{\rightsideA}{R_a}
\safemath{\rightsideB}{R_b}

\safemath{\indexS}{\setI_S} 

\safemath{\na}{n_a}			
\safemath{\nb}{n_b}			
\safemath{\coeffa}{p_i}	
\safemath{\coeffb}{q_j}	
\safemath{\seta}{\setP}		
\safemath{\setb}{\setQ}     
\safemath{\setw}{\setW}	
\safemath{\setz}{\setZ}	
\safemath{\cola}{\veca}		
\safemath{\colb}{\vecb}		
\safemath{\cold}{\vecd}		
\safemath{\inputvec}{\vecx} 	
\safemath{\error}{\vece}	
\safemath{\noiseout}{\vecz} 	
\safemath{\inputvecel}{x}
\safemath{\inputveca}{\vecx_a}
\safemath{\inputvecb}{\vecx_b}
\safemath{\outputvec}{\vecy}	
\safemath{\lambdamin}{\lambda_{\mathrm{min}}}

\safemath{\elltwo}{\ell_2}
\safemath{\ellone}{\ell_1}
\safemath{\ellzero}{\ell_0}
\safemath{\ellinf}{\ell_\infty}
\safemath{\ellinftilde}{\ell_{\widetilde\infty}}
\safemath{\licard}{Z(\coh,\coha,\cohb)}
\safemath{\xsol}{\hat{x}}
\safemath{\xbord}{x_b}		
\safemath{\xstat}{x_s}		
\safemath{\xstatLone}{\tilde{x}_s}
\safemath{\order}{\mathcal{O}} 
\safemath{\scales}{\Theta} 
\safemath{\ones}{\mathbf{1}} 
\safemath{\zeroes}{\mathbf{0}} 
\safemath{\thlone}{\kappa(\coh,\cohb)} 
\safemath{\constoneA}{\delta} 
\safemath{\constoneB}{\epsilon} 
\safemath{\nlarge}{L}				   
\safemath{\sumlarge}{S_\nlarge}
\safemath{\maxlarger}{P_\nlarge}	   
\safemath{\Pzero}{\textrm{P0}}	
\safemath{\Pone}{\textrm{P1}}
\safemath{\vecfir}{\vecw}			 
\safemath{\vecsec}{\vecz}
\safemath{\elvecfir}{w}              
\safemath{\elvecsec}{z}				 
\safemath{\nlargefir}{n}
\safemath{\normout}{\gamma}
\safemath{\auxfun}{h}
\safemath{\supp}{\textrm{supp}}

\safemath{\indexa}{\ell}
\safemath{\indexb}{r}
\safemath{\indexc}{i}
\safemath{\indexd}{j}

\safemath{\project}{P}

\newcommand{\dd}{\textnormal{d}}%

\usepackage{xcolor}

\newcommand{\sellr}[1]{\s_\ell^\text{R}#1}
\newcommand{\selli}[1]{\s_\ell^\text{I}#1}

\newcommand{\realpart}[1]{\textnormal{Re}\!\left\{ #1 \right\}\!}

\newcommand{\imagpart}[1]{\textnormal{Im}\!\left\{ #1 \right\}\!}

\safemath{\LAMA}{\textrm{LAMA}}

\safemath{\mLAMA}{\textrm{M-LAMA}}
\safemath{\smLAMA}{\textrm{SM-LAMA}}
\safemath{\mCBAMP}{\textrm{mcB-AMP}}

\safemath{\MRT}{\textrm{MRT}}
\safemath{\betamax}{\beta^\textnormal{max}}
\safemath{\tmax}{t_\textnormal{max}}
\safemath{\betamaxno}{\beta^\textnormal{max}}
\safemath{\betamin}{\beta^\textnormal{min}}
\safemath{\betaminno}{\beta^\textnormal{min}}

\safemath{\Nomin}{\No^\textnormal{min}(\beta)}
\safemath{\Nominnobeta}{\No^\textnormal{min}}
\safemath{\Nomax}{\No^\textnormal{max}(\beta)}
\safemath{\Nomaxnobeta}{\No^\textnormal{max}}

\safemath{\MAP}{\textrm{MAP}}
\safemath{\IO}{\textrm{IO}}
\safemath{\JO}{\textrm{JO}}
\safemath{\Nopost}{N_{0}^\textnormal{post}}
\safemath{\MT}{{M_\textnormal{T}}}
\safemath{\MR}{{M_\textnormal{R}}}
\safemath{\Tran}{\textnormal{T}}
\safemath{\Herm}{\textnormal{H}}
\safemath{\row}{\textnormal{r}}
\safemath{\col}{\textnormal{c}}

\IEEEoverridecommandlockouts 
\begin{document}

\title{On the Performance of Mismatched \\ Data Detection in Large MIMO Systems}

\author{Charles Jeon, Arian Maleki, and Christoph Studer
\thanks{C. Jeon and C.~Studer are with the School of ECE at Cornell University, Ithaca, NY; e-mails: {jeon@csl.cornell.edu}, {studer@cornell.edu}.}
\thanks{A. Maleki is with Department of Statistics at Columbia University, New York City, NY; e-mail: {arian@stat.columbia.edu}.}
\thanks{The work of CJ and CS was supported in part by Xilinx Inc.\ and by the US National Science Foundation under grants ECCS-1408006 and CCF-1535897.}
\thanks{An extended version of this paper including all proofs is in preparation \cite{JMS2015_journal}.}
}
\maketitle

\begin{abstract}
We investigate the performance of mismatched data detection in large multiple-input multiple-output (MIMO) systems, where the prior distribution of the transmit signal used in the data detector differs from the true prior.
To minimize the performance loss caused by this prior mismatch, we include a tuning stage into our recently-proposed large MIMO approximate message passing (LAMA) algorithm, which allows us to develop  mismatched LAMA  algorithms with optimal as well as sub-optimal tuning. 
We show that carefully-selected priors often enable simpler and computationally more efficient algorithms compared to LAMA with the true prior 
while achieving near-optimal performance. 
A performance analysis of our algorithms for a Gaussian prior and a uniform prior within a hypercube covering the QAM constellation recovers classical and recent results on linear and non-linear MIMO data detection,~respectively.
%

%
%
\end{abstract}


\section{Introduction}
\label{sec:intro}
Data detection in multiple-input multiple-output (MIMO) systems deals with the recovery of the transmit data vector $\bms_0\in\setO^\MT$, where $\setO$ is a finite constellation (e.g., QAM or PSK), from the noisy input-output relation \mbox{$\vecy=\bH\vecs_0+\bmn$}. In what follows, $\MT$ and $\MR$ denotes the number of transmit and receive antennas, respectively, $\bmy\in\complexset^\MR$ is the receive vector, \mbox{$\bH\in\complexset^{\MR\times\MT}$} is the MIMO system matrix, and \mbox{$\vecn\in\complexset^\MR$} is  i.i.d.\ circularly symmetric complex Gaussian noise.  To minimize the symbol-error rate, we are interested in solving the  individually-optimal (IO) data detection problem \cite{V1998,guo2003multiuser,GV2005}
\begin{align*}
(\text{IO})\quad
s_\ell^\text{IO} & = \argmax_{\tilde s_\ell\in\setO} \,p\!\left(\tilde s_\ell \,|\, \bmy, \bH\right)\!, \,\, \ell=1,\ldots,\MT.
\end{align*}
Here, $s_\ell^\text{IO}$ denotes the $\ell$-th IO estimate and $p\!\left(\tilde s_\ell \,|\, \bmy, \bH\right)$ is the conditional probability density function of  $\tilde s_\ell\in\setO$ given the receive vector $\bmy$ and the channel matrix $\bH$. 

The problem (IO) is known to be of combinatorial nature~\cite{V1998,guo2003multiuser,GV2005}, and the use of an exhaustive search or  sphere-decoding methods  results in prohibitive complexity for systems where~$\MT$ is large \cite{SJSB11}. 
In contrast, our recently proposed algorithm referred to as \underline{la}rge \underline{M}IMO \underline{a}pproximate message passing (LAMA)~\cite{JGMS2015conf}, achieves IO performance using a simple iterative procedure in the large-system limit, i.e., where we fix the system ratio $\beta=\MT/\MR$ and let $\MT\to\infty$. For finite-dimensional systems, LAMA was shown to deliver near-IO performance at low computational complexity~\cite{JGMS2015conf}. 
Despite all these advantages, LAMA requires repeated computations of transcendental functions that exhibit a high dynamic range. These computations render the design of corresponding high-throughput hardware designs that rely on finite precision (e.g., fixed-point) arithmetic a challenging task.

\subsection{Contributions}
We propose a mismatched version of the LAMA algorithm (short \mLAMA), which enables the design of hardware-friendly data detectors that achieve near-IO performance. 
%
We first develop a mismatched version of the complex Bayesian approximate message passing (cB-AMP) algorithm \cite{JGMS2015} that includes a tuning stage to minimize the performance loss caused by the mismatch in the signal prior.
We propose the mismatched state-evolution (SE) framework to enable a performance analysis in the large-system limit. 
We then apply our framework to mismatched data detection in large MIMO systems by considering two mismatched prior distributions: (i) a Gaussian prior and (ii) a uniform prior within a hypercube covering the QAM constellation.
We analyze the performance of the resulting mismatched algorithms in the large-system limit and demonstrate their efficacy in finite-dimensional systems.
%


\subsection{Relevant prior art}
%
%
The performance of zero forcing (ZF) and minimum mean-square error (MMSE) detection, which are both well-known instances of mismatched detection algorithms, has been investigated in \cite{VS1999,TH1999,SV2001} for the large-system limit.
The use of a uniform prior within a hypercube leads to an alternative mismatched 
detector for antipodal (e.g., BPSK) signals~\cite{tan2001constrained,yener2002cdma,TAXH2015}.
Corresponding theoretical results in \cite{DT2011,MR2011} for noiseless systems revealed that a system ratio of $\beta<2$ enables perfect signal recovery. The error-rate performance for the noisy case was derived recently in \cite{TAXH2015}.
The analysis of the mismatched algorithms presented in our paper recovers all these results.

The proposed \mLAMA algorithm 
relies upon approximate message passing (AMP) \cite{donoho2009,bayatimontanari,andreaGMCS}, 
developed for sparse signal recovery. 
The case of mismatched estimation of sparse signals via AMP was studied in~\cite{MMB2013,MMB2015}, where the performance of AMP was analyzed when the true prior is unknown.
This AMP algorithm includes a tuning stage for automated, optimal parameter selection that minimizes the output mean-squared error (MSE). 
%
%
The key differences between \mLAMA and the results in \cite{MMB2013,MMB2015} are that (i) we consider MIMO data detection and (ii) we know the true signal prior and intentionally select a mismatched prior in order to design hardware-friendly data detection algorithms that enable near-IO performance.

\subsection{Notation}
Lowercase and uppercase boldface letters designate vectors and matrices, respectively. We define the adjoint of matrix $\bH$ as $\bH^\Herm$ and use $\left\langle\cdot\right\rangle$ to abbreviate $\left\langle \bmx \right\rangle = \frac{1}{N}\sum_{k=1}^N x_k$. 
%
%
A multivariate complex-valued Gaussian probability density function (pdf) is denoted by $\setC\setN(\bmm,\bK)$, where~$\bmm$ is the mean vector and $\bK$ the covariance matrix. $\Exop_X\!\left[\cdot\right]$ and $\Varop_X\!\left[\cdot\right]$ denotes the mean and variance with respect to the 
random variable~$X$, respectively.


\section{Mismatched Complex Bayesian AMP}\label{sec:cB-AMP}

We start by presenting a mismatched version of the complex Bayesian approximate message passing (cB-AMP) algorithm~\cite{JGMS2015} (short \mCBAMP), which enables the use a different prior distribution $\tilde p(\tilde \bms)$ than the true signal prior $p(\bms_0)$. 
%
%

\subsection{The mismatched cB-AMP (\mCBAMP) algorithm}\label{sec:MCBAMP_alg}

Given an i.i.d. prior distribution $p(\bms_0)=\prod_{\ell=1}^N  p(s_{0\ell})$ of the true  signal $\bms_0$ and a mismatched prior $\tilde  p(\tilde \bms)=\prod_{\ell=1}^N  \tilde p(\tilde s_\ell)$,
the proposed \mCBAMP algorithm corresponds to 
\begin{align}
\label{eq:decouple_estimate}
\tilde\sigma^2_{t} &= \textstyle \frac{1}{\MR}\vecnorm{\bmr^{t}}_2^2,\\\label{eq:tau_opt}
\tau^{t} &= \argmin_{\tau\geq0} \,\Exop_{S_0,Z}\!\left[\abs{\mathsf{F}^\text{mm}(S_0+\tilde\sigma_t Z,\tau)-S_0}^2\right]\!,\\\label{eq:F_tau}
\bms^{t+1} &= \mathsf{F}^\text{mm}\!\left(\bms^{t} + \bH^\Herm \bmr^{t},\tau^t\right)\!,\\\nonumber
\bmr^{t+1} &= \bmy - \bH\bms^{t+1} + \beta \bmr^t \!\left\langle\mathsf{F'}^\text{mm}(\bms^t+\bH^\Herm\bmr^t,\tau^t)
\right\rangle\!,
\end{align}
which is carried out for $\tmax$ iterations $t=1,\ldots,\tmax$.
The algorithm is initialized by $s^1_\ell = \Exop_{S_0}[S_0]$ for all $\ell=1,\ldots,\MT$, where $S_0\sim p(s_0)$, $\bmr^1 = \bmy - \bH\bms^1$, and $\mathsf{F'}^\text{mm}$ is the derivative of $\mathsf{F}^\text{mm}$ taken by its first argument. 
The function
%
\begin{align}\label{eq:F}
\mathsf{F}^\text{mm}(s_\ell,\tau) 
&= \textstyle \Exop_{\tilde S}[\tilde S\vert s_\ell]
= \int_\complexset \tilde s p(\tilde s\vert s_\ell,\tau)\dd \tilde s, 
\end{align}
is the posterior mean with respect to the mismatched prior $\tilde p(\tilde s_\ell)$ and the variance parameter $\tau$, $p(\tilde s\vert s_\ell,\tau)=\frac{1}{C}p(s_\ell\vert\tilde s,\tau) \tilde p(\tilde s)$, $C$ is a normalization constant, and $p( s_\ell\vert\tilde s,\tau)\sim\setC\setN(\tilde s,\tau)$.
The functions $\mathsf{F}^\text{mm}(s_\ell,\tau)$ and $\mathsf{F'}^\text{mm}$ operate element-wise on vectors and the expectation in \fref{eq:tau_opt} is taken with respect to the true prior distribution of $S_0\sim p(s_0)$ and $Z\sim\setC\setN(0,1)$.

%

The \mCBAMP algorithm differs from the original cB-AMP algorithm derived in \cite{JGMS2015} by the additional steps \fref{eq:decouple_estimate} and \fref{eq:tau_opt}. 
In every iteration, step \fref{eq:decouple_estimate} estimates the so-called decoupled noise variance $\sigma^2_t$ (see \fref{sec:decoupling_section}) and step~\fref{eq:tau_opt} tunes the variance parameter $\tau^t$ based on the estimate $\sigma^2_t$.
The tuning stage \fref{eq:tau_opt} ensures that \mCBAMP converges to the solution that minimizes $\sigma^2_t$ for every iteration (see \fref{sec:optimal_tuning}).

\subsection{Decoupling property of AMP-based algorithms}\label{sec:decoupling_section}
We will frequently use the following definition. 

\begin{defi}
Assume a MIMO system with $\MT$ transmit and $\MR$ receive antennas, and let the entries of~$\bH$ be  i.i.d.\ $\setC\setN(0,1/\MR)$.
%
We define the \emph{large-system limit} by fixing the system ratio $\beta = \MT/\MR$ and letting $\MT\to\infty$. \label{def:largesystemlimit}
\end{defi}

As shown in \cite{JGMS2015,Maleki2010phd,bayatimontanari}, AMP-based algorithms  decouple the MIMO system into parallel AWGN channels in the large-system limit, i.e., the quantity $\bmz^t=\bms^t+\bH^\Herm\bmr^t$ can be expressed equivalently as $\bms_0+ \bmw^t$, where $\bmw^t\sim\setC\setN(0,\sigma_t^2\bI_\MT)$ and $\sigma_t^2$ is the decoupled noise variance.
A key property of AMP-based algorithms is that the decoupled noise variance $\sigma_t^2$ can be tracked \emph{exactly} by the state evolution (SE) framework.
We formalize the \emph{mismatched} SE framework for \mCBAMP in \fref{thm:SE}.
This result is a specific instance of the SE framework in \cite{bayatimontanari} and has been adapted to \mCBAMP.

\begin{thm} \label{thm:SE}
Assume the large-system limit and that $\mathsf{F}^\text{mm}$ is Lipschitz continuous. 
Then, the decoupled noise variance $\sigma_{t+1}^2$ after~$t$ \mCBAMP iterations  is given by the coupled recursion
\begin{align}
\label{eq:SE_gamma}
\gamma_{t}^2 &= \argmin_{\gamma^2\geq0}\Psi^\text{mm}(\sigma_t^2,\gamma^2),\\
\label{eq:SE_MSE}
\sigma_{t+1}^2 &= \No + \beta\Psi^\text{mm}(\sigma_t^2,\gamma_t^2),
\end{align}
which is initialized by $\sigma^2_1 = \No + \beta\Varop_{S_0}[S_0]$. Here, $S_0\sim p(s_0)$  and the MSE function is defined by $\Psi^\text{mm}(\sigma_t^2,\gamma_t^2)=\Exop_{S_0,Z}\!\left[\abs{\mathsf{F}^\text{mm}(S_0+\sigma_tZ,\gamma_t^2)-S_0}^2\right]$, where the expectation is taken with respect to the 
$S_0$ 
and $Z\sim\setC\setN(0,1)$. 
\end{thm}


%
If the true prior is identical to the mismatched prior, i.e., $p(\bms_0)=\tilde p(\tilde \bms)$, we can show that \mCBAMP delivers the same  decoupled noise variance as cB-AMP without prior mismatch. 
%
\begin{lem} If 
$p(\bms_0)=\tilde p(\tilde \bms)$, then 
 the decoupled noise variance
$\sigma^2_{t+1}$ of \mCBAMP is equivalent to $\sigma^2_{t+1}$ of cB-AMP \cite{JGMS2015}.\label{lem:lama_already_opt}
\end{lem}
The proof of \fref{lem:lama_already_opt} follows by noting that the MSE function $\Psi^\text{mm}(\sigma_t^2,\gamma^2)$ in \fref{eq:SE_MSE} is minimized by $\gamma^2=\sigma_t^2$ in \fref{eq:SE_gamma}, and reduces to the conditional variance \cite{GWSS2011}. Therefore, \fref{eq:SE_MSE} reduces to the conventional SE recursion of cB-AMP \cite{JGMS2015}.


\subsection{Optimal tuning of the variance parameter $\tau$}\label{sec:optimal_tuning}
%
%
Before we discuss the tuning stage \fref{eq:tau_opt} in detail, we formalize what we mean by optimal tuning of the variance parameter~$\tau^t$.
For all iterations $t=1,\ldots,\tmax$, our goal is to minimize the decoupled noise variance $\sigma_{\tmax+1}^2$ given by \fref{thm:SE}, as the smallest $\sigma_{\tmax+1}^2$ minimizes the error probability of our algorithm.
Hence, optimal tuning tries to identify a sequence of variance parameters $\{\tau^1,\ldots,\tau^{\tmax}\}$ so that \mCBAMP ultimately leads to the smallest $\sigma_{\tmax+1}^2$;  sub-optimal choices of $\tau^t$ either lead to a higher $\sigma_{\tmax+1}^2$ or cause slower convergence to the smallest $\sigma_{\tmax+1}^2$.
We will use the following definition~\cite{MMB2015}.
\begin{defi}\label{def:asymp_opt} Assume the large-system limit and denote the decoupled noise variance of \mCBAMP obtained from the sequence $\{\tau^1,\ldots,\tau^{\tmax}\}$ as $\sigma^2_{\tmax+1}(\tau^1,\ldots,\tau^{\tmax})$.
A sequence of parameters $\{\tau^1_\star,\ldots,\tau^{\tmax}_\star\}$ is \emph{optimally tuned} at iteration~$\tmax$, if and only if for all $\{\tau^1,\ldots,\tau^{\tmax}\}$ with $\tau^t\in [0,\infty)$ we have
\begin{align}\label{eq:optimal_tune_MSE}
\sigma^2_{\tmax+1}(\tau^1_\star,\ldots,\tau^{\tmax}_\star)
\leq 
\sigma^2_{\tmax+1}(\tau^1,\ldots,\tau^{\tmax}).
\end{align}
\end{defi}
%
%
We next show that the tuning stage in \fref{eq:tau_opt}, which is carried out at every iteration of \mCBAMP, achieves the smallest $\sigma_{\tmax+1}^2$, i.e., optimally tunes the variance parameters~$\tau^t$.
%
%
We omit the proof details as it follows closely that in \cite[Sec. 4.4]{MMB2015}.

\begin{thm}\label{thm:thm3_7_mbb} Suppose $\{\tau_\star^1,\ldots,\tau_\star^{\tmax}\}$ are optimally-tuned for iteration $\tmax$. Then, for any $t<\tmax$, 
$\{\tau_\star^1,\ldots,\tau_\star^{t}\}$ are also optimally-tuned for iteration $t$. 
Thus, one can obtain $\tmax$ optimally-tuned variance parameters by optimizing $\tau^1_\star$ at $t=1$, and then, proceeding iteratively by optimizing $\tau^t_\star$ until $t=\tmax$.
\end{thm}


In general, the exact value of the decoupled noise variance~$\sigma_t^2$ that is needed for the tuning stage in step \fref{eq:tau_opt} to select $\tau^t_\star$ is unknown.
We therefore use the estimate $\hat\sigma_t^2 = \frac{1}{\MR}\vecnorm{\bmr^t}^2$ in step~\fref{eq:decouple_estimate}, which was shown to converge to the true decoupled noise variance~$\sigma_t^2$ in the large-system limit \cite{andreaGMCS}.

\subsection{Decomposition of complex-valued systems}
We now briefly discuss properties of \mCBAMP in complex-valued systems that will be necessary for our analysis of \mCBAMP in MIMO systems. 
%
%
In particular, we show that for special constellations, the complex-valued set $\setO$ can be exactly characterized by a suitably-chosen real-valued set $\realpart{\setO}$.
\begin{defi}\label{def:sep_const}For all $s\in\setO$, express $s$ as $s = a+ib$, where $a \in\realpart{\setO}$, $b\in\imagpart{\setO}$. Then, the constellation $\setO$ is \emph{separable} if $p(s) = p(a)p(b)$ holds for all $s\in\setO$ and $\realpart{\setO}=\imagpart{\setO}$.
\end{defi}
An example of a separable constellation is $M^2$-QAM with equally likely transmit symbols. For such a separable set $\setO$, the following lemma (with proof in \cite{JGMS2015}) provides the equivalence of the complex-valued and real-valued SE framework.

\begin{lem}\label{lem:identical_separable}For a separable constellation set $\setO$, the mismatched SE recursion in \fref{thm:SE} can be expressed equivalently by:
\begin{align*}
\sigma_{t+1}^2 
&=
\No + 2\beta\min_{\gamma^2\geq0}\Exop_{S_0^\text{R},Z^\text{R}}
\!\left[
\!\left(
	\mathsf{F}_\text{R}^\text{mm}(S_0^\text{R}+\sigma_t Z^\text{R},\gamma^2)
 - S_0^\text{R}
\right)^2
\right]\!,
\end{align*}
where $\mathsf{F}_\text{R}^\text{mm}$ is posterior mean function with respect to the 
constellation set $\setO^\text{R}=\realpart{\setO}$, $S_0^\text{R}\in\setO^\text{R}$ and $Z^\text{R}\sim\setN(0,1/2)$. 
\end{lem}
\subsection{Fixed-point analysis}\label{sec:fp_betas}

While the performance of \mCBAMP at every iteration 
can be characterized by the mismatched SE recursion equations in \fref{thm:SE}, we are interested in analyzing the performance of \mCBAMP for $\tmax\to\infty$.
In this case, the mismatched SE in \fref{thm:SE} converges to the following fixed-point equation:
\begin{align}\label{eq:fpequation}
\sigma^2_\star = \No + \beta\min_{\gamma^2\geq0}\Psi^\text{mm}(\sigma^2_\star,\gamma^2) = \No + \beta\Psi_\star^\text{mm}(\sigma^2_\star).
\end{align}
Thus, as $\tmax\to\infty$, $\sigma_{t+1}^2$ in \fref{thm:SE} 
converges to~$\sigma_\star^2$ as given by \fref{eq:fpequation}. 
In general, if there are multiple fixed points, then \mCBAMP
 converges to the largest fixed-point, which ultimately leads to a higher probability of error than that of the smallest fixed-point solution.
To provide conditions that ensure a unique fixed-point solution to~\fref{eq:fpequation} (see  \fref{sec:box}), we use the following lemma (the proof is given in \cite{JGMS2015}).

%
\begin{lem}\label{lem:unique_fp_MRT}
Fix $p(\bms_0)$ and $\tilde p(\tilde \bms)$. The \emph{minimum recovery threshold} (MRT) $\betamin$ for \mLAMA is defined by:
\begin{align}\label{eq:betamin}
\betamin = \textstyle \min\limits_{\sigma^2\geq0}
\!\left(
\frac{\dd \Psi^\text{mm}_\star(\sigma^2)}{\dd \sigma^2}
\right)^{-1}.
\end{align}
For all system ratios $\beta<\betamin$ the fixed-point solution in \fref{eq:fpequation} is unique.
Furthermore, let $\sigma^2_\star=\argmin\limits_{\sigma^2\geq0}
\!\left(
\dd \Psi^\text{mm}_\star(\sigma^2)/\dd \sigma^2
\right)^{-1}$.
If for any other $\sigma^2\neq \sigma^2_\star$, \mbox{$
\betamin\dd \Psi^\text{mm}_\star(\sigma^2)/\dd \sigma^2<1$}, then \mLAMA also has a unique fixed point at $\beta=\betamin$.
\end{lem}

%


\section{Mismatched Data Detection with 
Tuning} \label{sec:box}

We now apply the mismatched cB-AMP framework to mismatched data detection in large MIMO systems, and refer to the algorithm as \underline{m}ismatched \underline{la}rge \underline{M}IMO \underline{A}MP (\mLAMA). 

In what follows, we assume that the true prior is taken from a discrete constellation set $\setO$ with equally likely symbols, i.e., $p(s_{0\ell})=\frac{1}{\abs{\setO}} \sum_{a\in\setO}\delta(s_{0\ell}-a)$, where $\abs{\setO}$ is the cardinality of the set $\setO$; we also define $\Exop_{S_0}[\abs{S_0}^2] = E_s$.
Note that in MIMO systems, the true signal prior is generally known.
Hence, it is natural to ask why one should consider a mismatched prior,
especially since the \LAMA algorithm \cite{JGMS2015conf} that uses the true prior minimizes the error probability.
To answer this question, consider the posterior mean function \fref{eq:F} of \LAMA \cite{JGMS2015conf}
\begin{align} \label{eq:sadlamafunction} \textstyle
\mathsf{F}(r,\tau) = \frac{\sum_{a\in\setO}a\exp\left(-\frac{1}{\tau}\abs{r-a}^2\right)}{\sum_{a\in\setO}\exp\left(-\frac{1}{\tau}\abs{r-a}^2\right)},
\end{align} 
whose calculation requires high arithmetic precision. 
In fact, even the use of double-precision floating-point arithmetic becomes numerically unstable for small values of $\tau$. 
Hence, the development of hardware designs for \LAMA that use finite-precision arithmetic is challenging. In contrast, suitably-chosen mismatched priors can lead to hardware-friendly data detectors.

\subsection{Optimally-tuned 
data detection with a Gaussian prior}\label{sec:gaus_prior}
We now derive an \mLAMA algorithm variant with a mismatched Gaussian prior. 
Without loss of generality, we assume a standard complex Gaussian distribution for the mismatched prior, i.e., $\tilde p(\tilde s_\ell)\sim\setC\setN(0,1)$ as the variance parameter $\tau^t$ will be scaled accordingly to $E_s$ in the tuning stage \fref{eq:tau_opt}.
For the mismatched Gaussian prior, the posterior mean function \fref{eq:F} is given by $\mathsf{F}^\text{mm}(r,\tau)= \frac{E_s}{E_s+\tau}r$. 
In the large-system limit, we can derive the following mismatched SE recursion given in \fref{eq:SE_MSE} using \fref{thm:SE}:
\begin{align}
\label{eq:gaussPsi}
\sigma_{t+1}^2 =  \textstyle \No + \beta
\min_{\gamma^2\geq0}
\frac{E_s}{(
E_s+\gamma^2)^2}(E_s\sigma_t^2+\gamma^4).
\end{align}
Evidently, the mismatched SE recursion \fref{eq:gaussPsi} only depends on the signal energy $E_s$ and no other properties of the true prior $p(\bms_0)$. 
This fact allows us to optimally tune the variance parameters only with knowledge of signal energy $E_s$.
We note that the RHS of \fref{eq:gaussPsi} is minimized by $\gamma^2=\sigma_t^2$ and thus, the mismatched SE recursion becomes:
%
%
\begin{align}
\sigma_{t+1}^2 = \textstyle \No + \beta\frac{E_s}{E_s+\sigma_t^2}\sigma_t^2. \label{eq:mmsefixedeq}
\end{align}
By \fref{lem:unique_fp_MRT}, \mLAMA has a unique fixed point if $\beta\leq1$.
%
Interestingly, if we define signal-to-interference ratio (SIR) as $\textit{SIR}=1/\sigma^2$ and let $\tmax\to\infty$, then the fixed-point solution of~\fref{eq:mmsefixedeq} coincides to the SIR of the  linear MMSE detector in the large-system limit \cite{TH1999,VS1999,SV2001}.
Hence, for a mismatched Gaussian prior, \mLAMA achieves the same performance as the linear MMSE detector.
We note that the proofs given in \cite{TH1999,VS1999,SV2001} use results from random matrix theory, whereas our analysis uses the mismatched SE recursion in \fref{thm:SE}. Furthermore, our result is constructive, i.e., \mLAMA is a novel, computationally efficient algorithm that implements linear MMSE detection. 
\subsection{Sub-optimal 
data detection with a Gaussian prior}\label{sec:ZFMF}
We can replace the optimal tuning stage of $\tau^t$ in \fref{eq:tau_opt} for the \mLAMA algorithm by a fixed variance parameter choice, which leads to a sub-optimal, mismatched algorithm, referred to as sub-optimal \mLAMA (short \smLAMA). 
We now show that this approach leads to other well-known linear data detectors.
In particular, we obtain the following mismatched SE recursions in the large-system limit for the following two choices of variance parameters $\gamma_t^2\to0$ and $\gamma_t^2\to\infty$ in \fref{eq:SE_gamma}:
\begin{align*}
(\text{ZF})\!\quad\!\sigma_{t+1}^2\!\! &= \No + \beta\!\lim_{\gamma_t^2\to0}\!\Psi^\text{mm}(\sigma_t^2,\gamma_t^2)\!=\!\No+\beta\sigma_t^2,\\
(\text{MF})\!\quad\!\sigma_{t+1}^2\!\! &= \No + \beta\!\!\lim_{\gamma_t^2\to\infty}\!\!\Psi^\text{mm}(\sigma_t^2,\gamma_t^2)\!=\!\No+\beta\Varop_{S_0}[S_0].
\end{align*}
%
By \fref{lem:unique_fp_MRT}, (ZF) and (MF) has a unique fixed point if $\beta<1$ and for any finite $\beta$, respectively.
%
If $\beta<1$, then the solution to the fixed-point equation of (ZF) and (MF) coincides exactly to the SIR given by ZF and MF detector in the large-system limit~\cite{TH1999,VS1999,SV2001}, respectively. Hence, by choosing specific predefined and sub-optimal variance parameters $\gamma_t^2$, \smLAMA can be used to perform ZF and MF data detection.
%



%
%

\subsection{Optimally-tuned 
data detection with a 
hypercube prior}\label{sec:softboxlama}
We now derive an \mLAMA algorithm variant using a mismatched uniform distribution within a hypercube around the true prior distribution of square constellations (e.g., QPSK and 16-QAM).
For example, for a QPSK system with equally likely symbols, we use a mismatched prior 
distributed uniformly in the interval $[-1,+1]$ for both the real and imaginary part. 
For this mismatched prior, we use \fref{lem:identical_separable} to compute 
the posterior mean function independently for the real and imaginary part; the posterior mean function $\mathsf{F}^\text{mm}$ is given by
\begin{align}\label{eq:F_box}
\mathsf{F}^\text{mm}(s_\ell,\tau) = \, &  \textstyle
s_\ell + \frac{\tau}{2}\!\left(
\nu_-(\sellr,\tau/2) + 
i \nu_-(\selli,\tau/2) 
\right),
\end{align}
where we use 
 \mbox{$\sellr=\realpart{s_\ell}$}, \mbox{$\selli=\imagpart{s_\ell}$}, and 
\begin{align*} \textstyle
\nu_-(s_\ell,\tau)=\frac{
e^{-\frac{1}{2\tau}(s_\ell+\alpha)^2}-
e^{-\frac{1}{2\tau}(s_\ell-\alpha)^2}
}{
\sqrt{2\pi\tau}\left(
\Phi\left(\frac{s_\ell+\alpha}{\sqrt{\tau}}\right)-\Phi\left(\frac{s_\ell-\alpha}{\sqrt{\tau}}\right)
\right)
},
\end{align*}
with $\Phi(x)\!=\!\int^x_{-\infty}\frac{1}{\sqrt{2\pi}}e^{-u^2/2}\dd u$. The mismatched SE recursion can be obtained from \fref{thm:SE} and evaluated numerically. 

There are two disadvantages of this algorithm: 
\begin{inparaenum}[(i)]
\item The computation of the posterior mean function \fref{eq:F_box} is not efficient from a hardware perspective as it involves transcendental functions.
In fact, computing $\nu_-(s_\ell,\tau)$ requires---similar to that of the optimal \LAMA algorithm~\fref{eq:sadlamafunction}---excessively high numerical precision;
\item the tuning stage \fref{eq:tau_opt} turns out to be non-trivial and requires numerical methods to find a minimum. Hence, corresponding hardware designs are impractical. 
\end{inparaenum}
%
%

\subsection{Sub-optimal 
data detection with a hypercube prior}\label{sec:sblama}

Analogously to the ZF detector in \fref{sec:ZFMF}, we can derive a sub-optimal variant of \mLAMA (\smLAMA) with the hypercube prior from \fref{sec:softboxlama}, where we replace the tuning stage in \fref{eq:tau_opt} by the fixed choice $\tau^t\to0$.
%

%
This choice leads to a much simpler algorithm compared to the optimally-tuned \mLAMA algorithm and enables a detailed performance analysis.
The posterior mean function reduces to
\begin{align*}
\lim\limits_{\tau\to0}\mathsf{F}^\text{mm}(s_\ell,\tau) =\, &\sellr+ \sign(\sellr)\min\!\left\{\alpha-\abs{\sellr},0\right\}\\
&+i
\left(
\selli+ \sign(\selli)\min\!\left\{\alpha-\abs{\selli},0\right\}\right),
\end{align*}
and thus, $\lim_{\tau\to0}\mathsf{F}^\text{mm}(s_\ell,\tau)$ and its derivative can be evaluated efficiently in hardware. 
%
%
%
%
%
Furthermore, by fixing $\tau\to0$, the tuning stages in~\fref{eq:decouple_estimate} and \fref{eq:tau_opt} are no longer required.
%


We now present conditions on the system ratio $\beta$ for which \smLAMA has a unique fixed point. The following \fref{lem:ERT_QAM}, with proof in \fref{app:ERT_QAM}, shows that the MRT of \smLAMA for $M^2$-QAM is given by $\betamin=(1-1/M)^{-1}$. 
\begin{lem}\label{lem:ERT_QAM} Assume a $M^2$-QAM prior for $S_0$ with equally likely symbols. Then, the MRT for
\smLAMA is given by $\betamin=(1-1/M)^{-1}$. 
Moreover, 
\smLAMA has a unique fixed point at $\beta=\betamin$ regardless of the noise variance $\No$.
\end{lem}

Using \fref{lem:ERT_QAM}, we obtain the same MRT for $M$-PAM constellations by \fref{lem:identical_separable}. We omit the proof and refer to~\cite{JGMS2015}.

\begin{cor} \smLAMA has the same MRT for $M^2$-QAM and $M$-PAM in a complex and real-valued system, respectively.
\end{cor}

%
%
%


We now show that this computationally-efficient \smLAMA variant achieves the same performance as a well-known relaxation of the maximum likelihood data detection problem \cite{tan2001constrained,yener2002cdma,TAXH2015}. 
The algorithm, which is known as box relaxation (BOX) detector, solves the following convex problem:
\begin{align} 
\hat\vecs = \argmin_{\tilde\bms\in\complexset^\MT} \, \|\bmy-\bH\tilde\bms\|_2 \quad \mathrm{subject\,\,to}\,\, \|\tilde\bms\|_\infty\leq1 \label{eq:boxproblem}
\end{align}
and slices the individual entries of $\hat\vecs$ onto the QPSK (or BSPK) constellations. The next result shows that  \smLAMA achieves the same error rate performance as the BOX detector for $\beta<2$, while providing a simple and computationally efficient algorithm. The proof is given in \fref{app:hassibi}.
\begin{lem} Assume $\beta<2$ and the large system limit. Then, for a complex-valued MIMO system with QPSK  (or a real-valued MIMO system with BPSK), \smLAMA achieves the same error-rate performance as the BOX algorithm in \eqref{eq:boxproblem}.\label{lem:hassibi}
\end{lem}

%

\begin{figure}[tp]
\centering
%
\includegraphics[width=0.95\columnwidth]{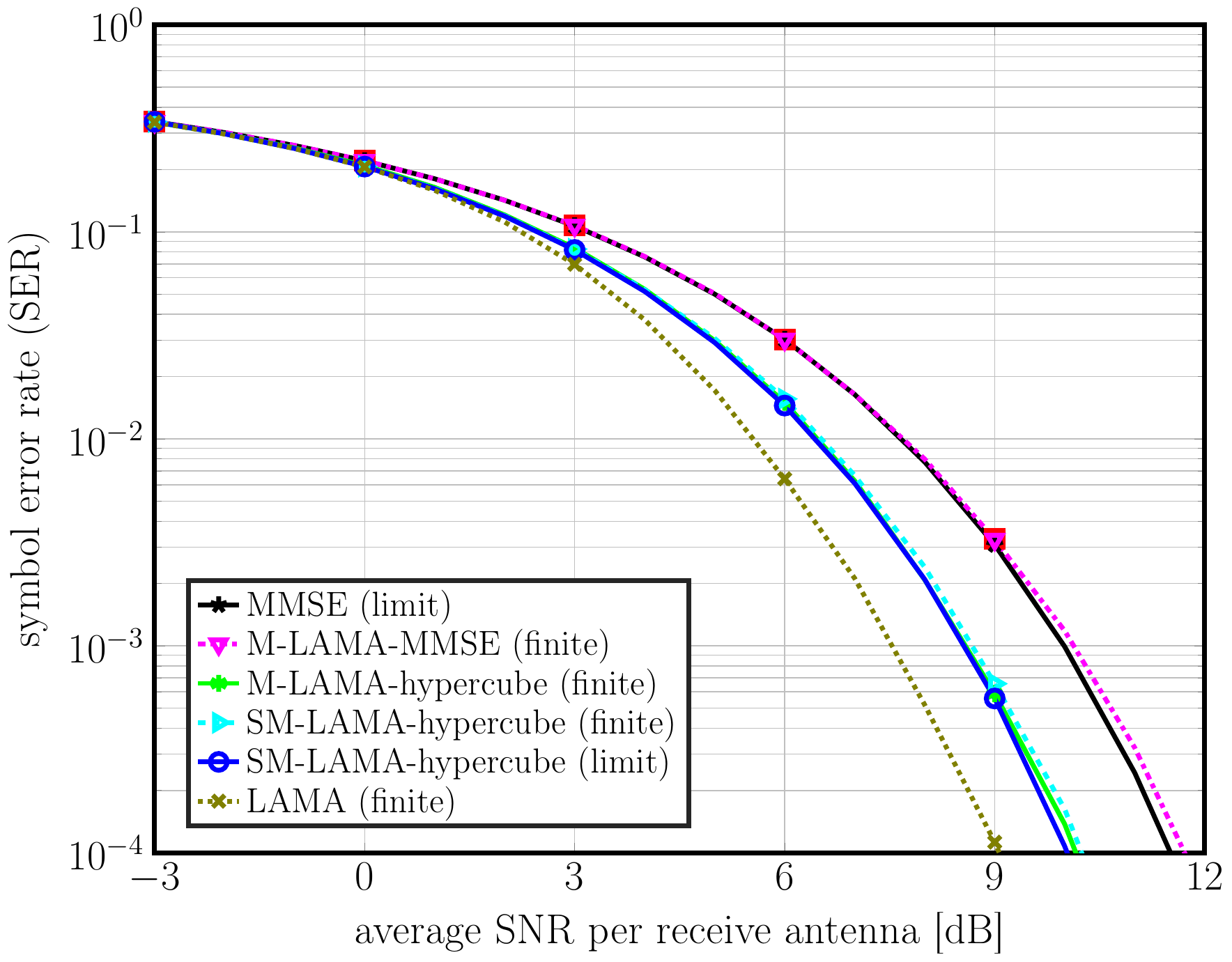}
\caption{
Symbol-error rate of \mLAMA and its variants for a $128\times64$ large-MIMO system with 10 iterations and QPSK;
(finite) and (limit) denote the detectors that are simulated and computed by SE (in the large-system limit), respectively.
\smLAMA with the uniform hypercube prior performs within 1\,dB
from LAMA \cite{JGMS2015conf} that achieves IO performance in the large-system~limit.
}
\label{fig:SE_QPSK}
\vspace{-0.5cm}
\end{figure}



\section{Numerical Results}\label{sec:numerical}

%
We now compare the error-rate performance of  the \mLAMA algorithm variants proposed in this paper.
While the mismatched SE framework in \fref{thm:SE} enables an exact performance analysis in the large-system limit, we also carry out Monte--Carlo simulations in a finite-dimensional large MIMO system with 128 receive and 64 transmit antennas.  
\fref{fig:SE_QPSK} shows the symbol error-rate performance of \LAMA, \mLAMA, and \smLAMA.
The optimally-tuned \mLAMA and sub-optimal \smLAMA with the uniform hypercube prior perform within 1\,dB of the LAMA algorithm~\cite{JGMS2015conf}, which achieves IO performance in the large-system limit. 
These results demonstrate that carefully-selected mismatched priors enable near-IO performance in finite-dimensional systems and in a hardware-friendly way.

\vspace{0.3cm}
\appendices

%


\section{Derivation of the MSE function $\Psi^\text{QAM}(\sigma^2)$}\label{app:qam_SE_deriv}

We will compute the MSE function for $\Psi^\text{QAM}(\sigma^2)$ by first computing the MSE function $\Psi^\text{PAM}(\sigma^2)$ for a real-valued $M$-PAM system with equally likely symbols and then, use \fref{lem:identical_separable} to express $\Psi^\text{QAM}(\sigma^2)$ for $M^2$-QAM.
We start by defining $\mathsf{F}^\alpha(s_\ell)=s_\ell+ \sign(s_\ell)\min\!\left\{\alpha-\abs{s_\ell},0\right\}$.
Note that for equally likely symbols, the $M$-PAM constellation can be expressed by $p(s_\ell)=\frac{1}{M}\sum_{k=-M/2+1}^{M/2}\delta(s_\ell-(2k-1))$.
Thus, by using symmetry for equally likely $M$-PAM symbols:
\begin{align}\label{eq:PAM_PSI}
&\Psi^\text{PAM}(\sigma^2) = \textstyle 
\frac{2}{M}\sum_{k=1}^{M/2}\Psi_k^\text{PAM}(\sigma^2),
\end{align}
where we introduced $\Psi^\text{PAM}_k(\sigma^2)$ that is defined by:
\begin{align*}
\Psi_k^\text{PAM}(\sigma^2)=& \,\textstyle \Exop_{Z}[(\mathsf{F}^\alpha((2k-1)+\sigma Z)-(2k-1))^2]\\
=&\, \textstyle\sigma^2 + (\bar\alpha_k ^2-\sigma^2)
Q\!\left(
\frac{\bar\alpha_k}{\sigma}
\right)+ (\alpha_k^2-\sigma^2)
Q\!\left(
\frac{\alpha_k }{\sigma}
\right)
\\&- \textstyle
\frac{\sigma}{\sqrt{2\pi}} \bar\alpha_k 
e^{
-\frac{\bar\alpha_k^2}{2\sigma^2}}
-
\frac{\sigma}{\sqrt{2\pi}}\alpha_k 
e^{
-\frac{\alpha_k ^2}{2\sigma^2}},
\end{align*}
with $\bar\alpha_k = \alpha-(2k-1)$, $\alpha_k = \alpha+(2k-1)$, $Z\sim\setN(0,1)$, and $Q(x)=\int_x^\infty \frac{1}{\sqrt{2\pi}} e^{-u^2/2}\dd u$.
%
By \fref{lem:identical_separable}, $\Psi^\text{QAM}$ can be computed
 by $\Psi^\text{QAM}(\sigma^2)=2\Psi^\text{PAM}(\sigma^2/2)$ with $\alpha=M-1$.

\section{Proof of \fref{lem:hassibi}}\label{app:hassibi}

We start by stating the following result from \cite[Thm. 2.1]{TAXH2015} that establishes the error-rate performance of BOX detector.
\begin{thm}[Theorem 2.1 \cite{TAXH2015}]\label{thm:hassibi} Assume a real-valued BPSK system with \mbox{$\beta<2$}. The bit-error rate in the large-system limit converges to $Q(1/\tau_\star)$, where $\tau_\star$ is the unique solution to $\tau_\star=\argmin_{\tau>0}g(\tau)$, where $g(\tau)$ is defined by
\begin{align*}
g(\tau)= \textstyle
\frac{\tau}{2}\!\left(\frac{1}{\beta}-\frac{1}{2}\right)\!
+\frac{\No}{2\beta\tau}
+
\frac{\tau}{2}
\int_{\frac{\tau}{2}}^\infty
\left(
x-\frac{2}{\tau}
\right)^2
\frac{1}{\sqrt{2\pi}}e^{-\frac{x^2}{2}}\dd x.
\end{align*}
\end{thm}

%
We note that as $\tau_\star$ is the unique, minimal solution to $g(\tau)$, we have $g'(\tau_\star)=0$.
%
%
%
%
%
Rearranging terms in $g'(\tau_\star)=0$ 
results in the fixed-point equation $\tau_\star^2 = \No + \beta\Psi(\tau_\star^2)$ with 
\begin{align}\label{eq:BPSK_hassibi_Fp}
\Psi(\tau_\star^2)= \textstyle
\frac{1}{2}\tau^2_\star - \frac{2\tau_\star }{\sqrt{2\pi}}e^{-\frac{2}{\tau_\star^2}}
+Q\!\left(\frac{2}{\tau_\star}\right)(4-\tau_\star^2).
\end{align}
The function $\Psi(\tau_\star^2)$ in \fref{eq:BPSK_hassibi_Fp} is identical to \fref{eq:PAM_PSI} for a BPSK system, i.e., $M=2$ with $\sigma^2=\tau_\star^2$.
This implies that the BOX-relaxed method in \cite{TAXH2015} and \smLAMA with the uniform hypercube prior achieves the same fixed-point \fref{eq:fpequation}. 
Moreover, due to the decoupling property of \smLAMA detailed in \fref{sec:decoupling_section}, the error-rate of a real-valued BPSK system in the large-system limit is given by $Q(1/\tau_\star)$.
The result of \fref{thm:hassibi} can be generalized to complex-valued QPSK systems and is equal to fixed-point solution of \smLAMA by \fref{lem:identical_separable}.

\section{Proof of \fref{lem:ERT_QAM}}\label{app:ERT_QAM}

As shown in \fref{app:qam_SE_deriv}, we compute $\frac{\dd }{\dd \sigma^2}\Psi^\text{QAM}(\sigma^2)$ and observe that   
$\frac{\dd }{\dd \sigma^2}\Psi^\text{QAM}(\sigma^2)\leq1-1/M$, where equality is achieved only when $\sigma^2\to0$. 
%
%
Thus, 
$\betamin=(1-1/M)^{-1}$.
%
%
To show that \mLAMA also has a unique fixed point when $\beta=\betamin$, we use \fref{lem:unique_fp_MRT} and 
observe that 
no other $\sigma^2_\star>0$ satisfies $\betamin = (\dd\Psi(\sigma^2)/\dd\sigma^2)^{-1}$ at $\sigma^2 = \sigma^2_\star$.

\vspace{0.17cm}

\balance

\bibliographystyle{IEEEtran}
\bibliography{bib/VIPabbrv,bib/confs-jrnls,bib/publishers,bib/VIP_160130}

\begin{thebibliography}{10}
\providecommand{\url}[1]{#1}
\csname url@samestyle\endcsname
\providecommand{\newblock}{\relax}
\providecommand{\bibinfo}[2]{#2}
\providecommand{\BIBentrySTDinterwordspacing}{\spaceskip=0pt\relax}
\providecommand{\BIBentryALTinterwordstretchfactor}{4}
\providecommand{\BIBentryALTinterwordspacing}{\spaceskip=\fontdimen2\font plus
\BIBentryALTinterwordstretchfactor\fontdimen3\font minus
  \fontdimen4\font\relax}
\providecommand{\BIBforeignlanguage}[2]{{%
\expandafter\ifx\csname l@#1\endcsname\relax
\typeout{** WARNING: IEEEtran.bst: No hyphenation pattern has been}%
\typeout{** loaded for the language `#1'. Using the pattern for}%
\typeout{** the default language instead.}%
\else
\language=\csname l@#1\endcsname
\fi
#2}}
\providecommand{\BIBdecl}{\relax}
\BIBdecl

\bibitem{JMS2015_journal}
C.~Jeon, A.~Maleki, and C.~Studer, ``Mismatched data detection in large {MIMO}
  systems,'' \emph{in preparation}.

\bibitem{V1998}
S.~Verd\'u, \emph{Multiuser Detection}.\hskip 1em plus 0.5em minus 0.4em\relax
  Cambridge University Press, 1998.

\bibitem{guo2003multiuser}
D.~Guo and S.~Verd\'u, ``Multiuser detection and statistical mechanics,'' in
  \emph{Commun., Inf. and Netw. Security}.\hskip 1em plus 0.5em minus
  0.4em\relax Springer, 2003, pp. 229--277.

\bibitem{GV2005}
------, ``Randomly spread {CDMA}: {Asymptotics} via statistical physics,''
  \emph{{IEEE} Trans. Inf. Theory}, vol.~51, no.~6, pp. 1983--2010, Jun. 2005.

\bibitem{SJSB11}
D.~Seethaler, J.~Jald{\'e}n, C.~Studer, and H.~B\"olcskei, ``On the complexity
  distribution of sphere decoding,'' \emph{{IEEE} Trans. Inf. Theory}, vol.~57,
  no.~9, pp. 5754--5768, Sept. 2011.

\bibitem{JGMS2015conf}
C.~Jeon, R.~Ghods, A.~Maleki, and C.~Studer, ``Optimality of large {MIMO}
  detection via approximate message passing,'' in \emph{Proc. IEEE Int. Symp.
  Inf. Theory (ISIT)}, Jun. 2015, pp. 1227--1231.

\bibitem{JGMS2015}
------, ``Optimal data detection in large {MIMO},'' \emph{in preparation}.

\bibitem{VS1999}
S.~Verd\'u and S.~Shamai, ``Spectral efficiency of {CDMA} with random
  spreading,'' \emph{{IEEE} Trans. Inf. Theory}, vol.~45, no.~2, pp. 622--640,
  Mar. 1999.

\bibitem{TH1999}
D.~Tse and S.~Hanly, ``Linear multiuser receivers: effective interference,
  effective bandwidth and user capacity,'' \emph{{IEEE} Trans. Inf. Theory},
  vol.~45, no.~2, pp. 641--657, Mar. 1999.

\bibitem{SV2001}
S.~Shamai and S.~Verdu, ``The impact of frequency-flat fading on the spectral
  efficiency of {CDMA},'' \emph{{IEEE} Trans. Inf. Theory}, vol.~47, no.~4, pp.
  1302--1327, May 2001.

\bibitem{tan2001constrained}
P.~H. Tan, L.~K. Rasmussen, and T.~J. Lim, ``Constrained maximum-likelihood
  detection in {CDMA},'' \emph{{IEEE} Trans. Commun.}, vol.~49, no.~1, pp.
  142--153, Jan. 2001.

\bibitem{yener2002cdma}
A.~Yener, R.~D. Yates, and S.~Ulukus, ``{CDMA} multiuser detection: A nonlinear
  programming approach,'' \emph{{IEEE} Trans. Commun.}, vol.~50, no.~6, pp.
  1016--1024, Jun. 2002.

\bibitem{TAXH2015}
C.~Thrampoulidis, E.~Abbasi, W.~Xu, and B.~Hassibi, ``{BER} analysis of the box
  relaxation for {BPSK} signal recovery,'' \emph{IEEE Int. Conf. Acoust.,
  Speech, Signal Process. (ICASSP)}, 2016.

\bibitem{DT2011}
D.~L. Donoho and J.~Tanner, ``Counting the faces of randomly-projected
  hypercubes and orthants, with applications,'' \emph{Discrete Comput.
  Geometry}, vol.~43, no.~3, pp. 522--541, Apr. 2010.

\bibitem{MR2011}
O.~Mangasarian and B.~Recht, ``Probability of unique integer solution to a
  system of linear equations,'' \emph{Eur. J. Oper. Res.}, vol. 214, no.~1, pp.
  27--30, Oct. 2011.

\bibitem{donoho2009}
D.~L. Donoho, A.~Maleki, and A.~Montanari, ``Message-passing algorithms for
  compressed sensing,'' \emph{Proc. Natl. Acad. Sci. USA}, vol. 106, no.~45,
  pp. 18\,914--18\,919, Nov. 2009.

\bibitem{bayatimontanari}
M.~Bayati and A.~Montanari, ``The dynamics of message passing on dense graphs,
  with applications to compressed sensing,'' \emph{{IEEE} Trans. Inf. Theory},
  vol.~57, no.~2, pp. 764--785, Feb. 2011.

\bibitem{andreaGMCS}
A.~Montanari, \emph{Graphical models concepts in compressed sensing, Compressed
  Sensing (Y.C. Eldar and G. Kutyniok, eds.)}.\hskip 1em plus 0.5em minus
  0.4em\relax Cambridge University Press, 2012.

\bibitem{MMB2013}
A.~{Mousavi}, A.~{Maleki}, and R.~G. {Baraniuk}, ``{Parameterless Optimal
  Approximate Message Passing},'' \emph{arXiv:1311.0035 [cs.IT]}, Oct. 2013.

\bibitem{MMB2015}
------, ``{Consistent Parameter Estimation for {LASSO} and Approximate Message
  Passing},'' \emph{arXiv:1511.01017 [math.ST]}, Nov. 2015.

\bibitem{Maleki2010phd}
A.~Maleki, ``Approximate message passing algorithms for compressed sensing,''
  Ph.D. dissertation, Stanford University, Jan. 2011.

\bibitem{GWSS2011}
D.~Guo, Y.~Wu, S.~Shamai, and S.~Verd\'u, ``Estimation in {Gaussian} noise:
  Properties of the minimum mean-square error,'' \emph{{IEEE} Trans. Inf.
  Theory}, vol.~57, no.~4, pp. 2371--2385, Apr. 2011.

\end{thebibliography}

\end{document}